\documentclass[12pt,prd,showpacs,tightenlines,nofootinbib]{revtex4}
\usepackage{bm}
\usepackage{graphics}
\usepackage{rotating}
\usepackage{epsfig}
\begin{document}
\begin{flushright}{HU-EP-06/42}\end{flushright}
\title{New analysis of semileptonic $B$ decays in
the relativistic quark model} 
\author{D. Ebert}
\affiliation{Institut f\"ur Physik, Humboldt--Universit\"at zu Berlin,
Newtonstr. 15, D-12489  Berlin, Germany}
\author{R. N. Faustov}
\author{V. O. Galkin}
\affiliation{Institut f\"ur Physik, Humboldt--Universit\"at zu Berlin,
Newtonstr. 15, D-12489 Berlin, Germany}
\affiliation{Dorodnicyn Computing Centre, Russian Academy of Sciences,
  Vavilov Str. 40, 119991 Moscow, Russia}

\begin{abstract}
We present the new analysis of the semileptonic $B$ decays
in the framework of the relativistic quark model based on the
quasipotential approach. Decays both to heavy $D^{(*)}$ and light
$\pi(\rho)$ mesons are considered. All relativistic effects are
systematically taken into account including contributions of the
negative-energy states and the wave function transformation from the rest
to moving reference frame. For heavy-to-heavy transitions the
heavy quark expansion is applied. Leading and subleading Isgur-Wise
functions are determined as the overlap integrals of initial and final
meson wave functions. For heavy-to-light transitions the explicit
relativistic expressions are used to determine the dependence of the
form factors on the momentum transfer squared. Such treatment significantly
reduces theoretical uncertainties and increases reliability of
obtained predictions. All results for form factors, partial and total
decay rates agree well with recent experimental data and unquenched
lattice calculations. From this comparison we find the following
values of the Cabibbo-Kobayashi-Maskawa  matrix elements:
$|V_{cb}|=(3.85\pm0.15\pm 0.20)\times 10^{-2}$ and
 $|V_{ub}|=(3.82\pm0.20\pm0.20)\times 10^{-3}$, where the first error
 is experimental and the second one is theoretical.

\end{abstract}


\maketitle

\section{Introduction}
\label{intr}
The investigation of the semileptonic decays of heavy $B$ mesons is an
important source for the determination of the parameters of the standard model,
such as Cabibbo-Kobayashi-Maskawa (CKM) matrix elements $V_{cb}$ and
$V_{ub}$. They also provide valuable insight in quark dynamics in the
nonperturbative domain of QCD. Recently significant experimental
progress has been achieved in studying exclusive semileptonic $B$
decays (for a recent review see, e.g., Ref.~\cite{kowal}). A very
important information on both the values of the total decay rates and 
the differential decay rate dependence on the momentum transfer
is becoming available. The experimental accuracy is constantly
increasing due to the large data accumulated at $B$ factories. For
most decay modes experimental errors are comparable to or already smaller
than theoretical ones. Thus the accurate extraction of the CKM matrix
elements from the semileptonic $B$ decays requires an increase in the
reliability and 
precision of theoretical methods for determining weak decay form
factors. Main theoretical approaches for calculating these form
factors are lattice QCD, QCD sum rules and constituent quark
models. Unfortunately, the first two approaches are applicable only in
limited regions: lattice QCD gives reliable results for high values of
the square of momentum transfer from the parent to daughter hadron
$q^2$, while QCD sum rules are suitable only for low $q^2$. Thus
extrapolations for the full $q^2$ region are at present inevitable in
these methods. On the other hand, quark models, which are not straightforwardly
related to QCD (at least this relation is not currently
established), allow to determine weak form factors in the whole
kinematical range. To give a correct description of semileptonic decays
such models should consistently and comprehensively account for
relativistic effects and use the wave functions which lead to correct
meson masses.  They should also respect all the relations imposed by
the symmetries of the QCD Lagrangian arising in the heavy quark limit.              

From a theoretical point of view the simplest semileptonic $B$ decays
are heavy-to-heavy transitions such as $B\to D^{(*)}e\nu$. The presence of
the heavy quark both in the initial and final mesons significantly
simplifies the understanding of such processes. The heavy quark
limit $m_Q\to\infty$ is a good initial approximation
 \cite{iw}. In this limit heavy quark symmetry emerges 
substantially reducing the number of independent characteristics which
are necessary  for the description of heavy-light meson properties
\cite{iw1}. Mass and spin decouple from the consideration and all
meson properties are determined by the light quark degrees of freedom
alone. This leads to symmetry relations between form
factors responsible for the heavy-to-heavy weak transitions. Thus one
needs considerably less independent functions. For the semileptonic
$B$ decay to the ground state $D^{(*)}$ meson all form factors
can be expressed through one Isgur-Wise function \cite{iw} which
is normalized to 1 at the point of zero recoil of the final
meson. However, in reality $b$ and $c$ quarks are not infinitely heavy
and therefore the corrections in inverse powers of the heavy quark
mass $m_Q$ (especially $m_c$) are important. Heavy quark effective
theory (HQET) \cite{n} is the adopted tool for a systematic
expansion of weak decay amplitudes in  $1/m_Q$. The coefficients in this
expansion are functions of the velocity transfer in the weak decay and
do not depend on spin and flavour of the heavy quarks. These functions
originate from the infrared (nonperturbative) region of QCD and thus
cannot be determined from first principles of QCD at present. It is
necessary to use some model assumptions in order to calculate these
functions.  Note that notwithstanding heavy quark
symmetry violation already in the first order in  $1/m_Q$, its
remnants remain and thus HQET significantly restricts the structure of
$1/m_Q$ corrections.  E.g., in the first order in $1/m_Q$ for $B$
decays to the ground state $D^{(*)}$ mesons one mass parameter and
four additional functions (two of which are normalized at the zero
recoil point) emerge instead of 12 possible ones. This is the
consequence of QCD and heavy quark symmetry. Thus all models which
pretend to describe correctly weak heavy-to-heavy transitions should
satisfy HQET symmetry relations. 

The theoretical description of exclusive semileptonic heavy-to-light $B$
decays, such as $B\to\pi(\rho)e\nu$, represents a more difficult task
since the final meson contains light quarks only. The expansion in
inverse powers of the $b$ quark mass does
not reduce the number of independent form
factors. Only relations between semileptonic and rare radiative decays
emerge in the heavy $b$ quark limit. It is important to note that in
these decays the final light meson has a large (compared to its mass)
recoil momentum (energy) in the rest frame of the decaying $B$ meson almost in the
whole kinematical range except the small region near the point of zero
recoil. The maximum value of recoil momentum is of order of
$m_b/2$. Thus near this point one can expand both in inverse powers of
the heavy $b$ quark mass and large recoil momentum of the final light
meson. Such expansions lead to the so-called large-energy effective
theory (LEET) \cite{dg}, to a new symmetry and as a result to the form factor
relations in the heavy 
quark and large recoil limits \cite{clopr,efgfr}. Large values of the recoil
momentum require the completely relativistic treatment of the
semileptonic heavy-to-light $B$ decays.

In our previous papers  \cite{fg,fgmslhh,fgmslhl} we considered
semileptonic $B$ decays to heavy $D^{(*)}$ and light $\pi(\rho)$
mesons in the framework of the relativistic quark model based on the
quasipotential approach. At that time taking into account large
experimental errors in the measured decay rates we used the simple
Gaussian parameterization for the meson wave functions. Moreover, we
calculated the heavy-to-light decay form factors only at the point of
maximum recoil of the final light meson using an expansion both in
inverse powers of the heavy $b$ quark and large recoil momentum of the
light meson. Then we employed a Gaussian or pole parameterization of the
form factors in order to extrapolate them to the whole kinematical
range. Such substitutions and extrapolations induced
significant theoretical errors in the obtained results, but the
accuracy was sufficient compared to large experimental uncertainties.
Since then, as it has been already mentioned before, the experimental
accuracy improved significantly. The main aim of this paper is to
revise our previous considerations of semileptonic $B$ decays
substantially increasing the precision and reliability. This is achieved
by using the wave functions of the heavy $B$, $D$ and light $\pi$,
$\rho$ mesons which were obtained by calculating their mass spectra
\cite{hlm,lmm}.  The complete expressions for the heavy-to-light decay
form factors are used for calculations and the determination of their
$q^2$ dependence, thus avoiding ad hoc parameterizations.      
In the following we concentrate on the study of the relativistic
effects and, for simplicity, neglect short-distance radiative corrections
\cite{n}  since their contribution does not exceed the uncertainty
of our calculations.    

The paper is organized as follows. In Sec.~\ref{rqm} we briefly
describe our relativistic quark model. Then in Sec.~\ref{mml}   we
discuss the relativistic calculation of the decay matrix element of
the weak current between meson states in the quasipotential
approach. Special attention is devoted to the contributions of the
negative energy states and the relativistic transformation of the wave
functions from the rest to the moving reference frame. Semileptonic $B$
decays to $D^{(*)}$ mesons are considered in Sec.~\ref{sec:bdsl} using
the heavy quark expansion in $1/m_Q$. Leading and subleading Isgur-Wise
functions are explicitly determined as the overlap integrals of the
initial and final meson wave functions. A comprehensive comparison with
recent experimental data is given and on this basis the value of the
CKM matrix element $|V_{cb}|$ is determined.  In Sec.~\ref{sec:blsd}
semileptonic $B$ decays to $\pi$ and $\rho$ are investigated. The
parameterization of the calculated form factors in the whole
kinematical range is given. Total and partial decay rates are compared
with recent measurements and the value of the CKM matrix element
$|V_{ub}|$ is extracted. Section~\ref{sec:concl} contains our conclusions.

\section{Relativistic quark model}  
\label{rqm}

In the quasipotential approach a meson is described by the wave
function of the bound quark-antiquark state, which satisfies the
quasipotential equation \cite{3} of the Schr\"odinger type \cite{4}
\begin{equation}
\label{quas}
{\left(\frac{b^2(M)}{2\mu_{R}}-\frac{{\bf
p}^2}{2\mu_{R}}\right)\Psi_{M}({\bf p})} =\int\frac{d^3 q}{(2\pi)^3}
 V({\bf p,q};M)\Psi_{M}({\bf q}),
\end{equation}
where the relativistic reduced mass is
\begin{equation}
\mu_{R}=\frac{E_1E_2}{E_1+E_2}=\frac{M^4-(m^2_1-m^2_2)^2}{4M^3},
\end{equation}
and $E_1$, $E_2$ are the center of mass energies on mass shell given by
\begin{equation}
\label{ee}
E_1=\frac{M^2-m_2^2+m_1^2}{2M}, \quad E_2=\frac{M^2-m_1^2+m_2^2}{2M}.
\end{equation}
Here $M=E_1+E_2$ is the meson mass, $m_{1,2}$ are the quark masses,
and ${\bf p}$ is their relative momentum.  
In the center of mass system the relative momentum squared on mass shell 
reads
\begin{equation}
{b^2(M) }
=\frac{[M^2-(m_1+m_2)^2][M^2-(m_1-m_2)^2]}{4M^2}.
\end{equation}

The kernel 
$V({\bf p,q};M)$ in Eq.~(\ref{quas}) is the quasipotential operator of
the quark-antiquark interaction. It is constructed with the help of the
off-mass-shell scattering amplitude, projected onto the positive
energy states. 
Constructing the quasipotential of the quark-antiquark interaction, 
we have assumed that the effective
interaction is the sum of the usual one-gluon exchange term with the mixture
of long-range vector and scalar linear confining potentials, where
the vector confining potential
contains the Pauli interaction. The quasipotential is then defined by
\cite{mass1}
  \begin{equation}
\label{qpot}
V({\bf p,q};M)=\bar{u}_1(p)\bar{u}_2(-p){\mathcal V}({\bf p}, {\bf
q};M)u_1(q)u_2(-q),
\end{equation}
with
$${\mathcal V}({\bf p},{\bf q};M)=\frac{4}{3}\alpha_sD_{ \mu\nu}({\bf
k})\gamma_1^{\mu}\gamma_2^{\nu}
+V^V_{\rm conf}({\bf k})\Gamma_1^{\mu}
\Gamma_{2;\mu}+V^S_{\rm conf}({\bf k}),$$
where $\alpha_s$ is the QCD coupling constant, $D_{\mu\nu}$ is the
gluon propagator in the Coulomb gauge
\begin{equation}
D^{00}({\bf k})=-\frac{4\pi}{{\bf k}^2}, \quad D^{ij}({\bf k})=
-\frac{4\pi}{k^2}\left(\delta^{ij}-\frac{k^ik^j}{{\bf k}^2}\right),
\quad D^{0i}=D^{i0}=0,
\end{equation}
and ${\bf k=p-q}$; $\gamma_{\mu}$ and $u(p)$ are 
the Dirac matrices and spinors
\begin{equation}
\label{spinor}
u^\lambda({p})=\sqrt{\frac{\epsilon(p)+m}{2\epsilon(p)}}
\left(
\begin{array}{c}1\cr {\displaystyle\frac{\bm{\sigma}
      {\bf  p}}{\epsilon(p)+m}}
\end{array}\right)\chi^\lambda.
\end{equation}
Here  $\bm{\sigma}$   and $\chi^\lambda$
are the Pauli matrices and spinors; $\epsilon(p)=\sqrt{p^2+m^2}$.
The effective long-range vector vertex is
given by
\begin{equation}
\label{kappa}
\Gamma_{\mu}({\bf k})=\gamma_{\mu}+
\frac{i\kappa}{2m}\sigma_{\mu\nu}k^{\nu},
\end{equation}
where $\kappa$ is the Pauli interaction constant characterizing the
long-range anomalous chromomagnetic moment of quarks. Vector and
scalar confining potentials in the nonrelativistic limit reduce to
\begin{eqnarray}
\label{vlin}
V_V(r)&=&(1-\varepsilon)(Ar+B),\nonumber\\ 
V_S(r)& =&\varepsilon (Ar+B),
\end{eqnarray}
reproducing 
\begin{equation}
\label{nr}
V_{\rm conf}(r)=V_S(r)+V_V(r)=Ar+B,
\end{equation}
where $\varepsilon$ is the mixing coefficient. 

The expression for the quasipotential of the heavy quarkonia,
expanded in $v^2/c^2$ without and with retardation corrections to the
confining potential, can be found in Refs.~\cite{mass1,mass},
respectively. The 
structure of the spin-dependent interaction is in agreement with
the parameterization of Eichten and Feinberg \cite{ef}. The
quasipotential for the heavy quark interaction with a light antiquark
without employing the nonrelativistic ($v/c$)  expansion for the light quark
is given in Ref.~\cite{hlm}.  All the parameters of
our model like quark masses, parameters of the linear confining potential
$A$ and $B$, mixing coefficient $\varepsilon$ and anomalous
chromomagnetic quark moment $\kappa$ are fixed from the analysis of
heavy quarkonium masses \cite{mass1} and radiative
decays \cite{gf}. The quark masses
$m_b=4.88$ GeV, $m_c=1.55$ GeV, $m_{u,d}=0.33$ GeV and
the parameters of the linear potential $A=0.18$ GeV$^2$ and $B=-0.30$ GeV
have usual values of quark models.  The value of the mixing
coefficient of vector and scalar confining potentials $\varepsilon=-1$
has been determined from the consideration of the heavy quark expansion
for the semileptonic $B\to D$ decays
\cite{fg} and charmonium radiative decays \cite{gf}.
Finally, the universal Pauli interaction constant $\kappa=-1$ has been
fixed from the analysis of the fine splitting of heavy quarkonia ${
}^3P_J$- states \cite{mass1}. Note that the 
long-range  magnetic contribution to the potential in our model
is proportional to $(1+\kappa)$ and thus vanishes for the 
chosen value of $\kappa=-1$. It has been known for a long time 
that the correct reproduction of the
spin-dependent part of the quark-antiquark interaction requires 
either assuming  the scalar confinement or equivalently  introducing the
Pauli interaction with $\kappa=-1$ \cite{schn,mass1,mass} in the vector
confinement.

\section{Matrix elements of the weak current for
  $\bm{\lowercase{b\to c,u}}$ transitions} \label{mml}

In order to calculate the exclusive semileptonic decay rate of the
$B$ meson, it is necessary to determine the corresponding matrix
element of the  weak current between meson states.
In the quasipotential approach,  the matrix element of the weak current
$J^W_\mu=\bar f\gamma_\mu(1-\gamma_5)b$, associated with $b\to f$ ($f=c$
or $u$) transition, between a $B$ meson with mass $M_{B}$ and
momentum $p_{B}$ and a final meson $F$ ($F=D^{(*)}$ or $\pi(\rho)$)
with mass $M_F$  and momentum $p_F$ takes the form \cite{f} 
\begin{equation}\label{mxet} 
\langle F(p_F) \vert J^W_\mu \vert B(p_{B})\rangle
=\int \frac{d^3p\, d^3q}{(2\pi )^6} \bar \Psi_{F\,{\bf p}_F}({\bf
p})\Gamma _\mu ({\bf p},{\bf q})\Psi_{B\,{\bf p}_{B}}({\bf q}),
\end{equation}
where $\Gamma _\mu ({\bf p},{\bf
q})$ is the two-particle vertex function and  
$\Psi_{M\,{\bf p}_M}$ are the
meson ($M=B,F)$ wave functions projected onto the positive energy states of
quarks and boosted to the moving reference frame with momentum ${\bf p}_M$.
\begin{figure}
  \centering
  \includegraphics{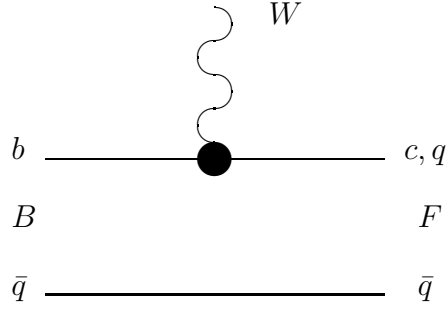}
\caption{Lowest order vertex function $\Gamma^{(1)}$
contributing to the current matrix element (\ref{mxet}). \label{d1}}
\end{figure}

\begin{figure}
  \centering
  \includegraphics{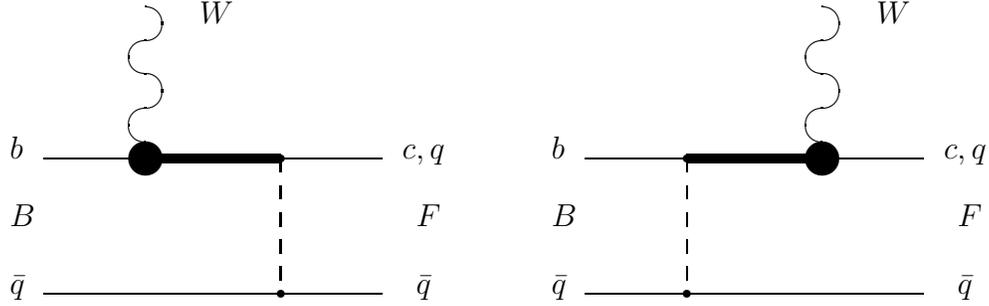}
\caption{ Vertex function $\Gamma^{(2)}$
taking the quark interaction into account. Dashed lines correspond  
to the effective potential ${\cal V}$ in 
(\ref{qpot}). Bold lines denote the negative-energy part of the quark
propagator. \label{d2}}
\end{figure}

 The contributions to $\Gamma$ come from Figs.~\ref{d1} and \ref{d2}. 
The leading order vertex function $\Gamma^{(1)}$ corresponds to the impulse
approximation, while the vertex function $\Gamma^{(2)}$ accounts for
contributions of the negative-energy states. Note that the form of the
relativistic corrections resulting from the vertex function
$\Gamma^{(2)}$ is explicitly dependent on the Lorentz structure of the
quark-antiquark interaction. In the leading order of the  the heavy
quark expansion  ($m_{b,c}\to \infty$) for $B\to D$ transitions 
only $\Gamma^{(1)}$ contributes, while $\Gamma^{(2)}$  
contributes already at the subleading order. 
The vertex functions look like
\begin{equation} \label{gamma1}
\Gamma_\mu^{(1)}({\bf
p},{\bf q})=\bar u_{f}(p_f)\gamma_\mu(1-\gamma^5)u_b(q_b)
(2\pi)^3\delta({\bf p}_q-{\bf
q}_q),\end{equation}
and
\begin{eqnarray}\label{gamma2} 
\Gamma_\mu^{(2)}({\bf
p},{\bf q})&=&\bar u_{f}(p_f)\bar u_q(p_q) \Bigl\{\gamma_{1\mu}(1-\gamma_1^5)
\frac{\Lambda_b^{(-)}(
k)}{\epsilon_b(k)+\epsilon_b(p_q)}\gamma_1^0
{\cal V}({\bf p}_q-{\bf
q}_q)\nonumber \\ 
& &+{\cal V}({\bf p}_q-{\bf
q}_q)\frac{\Lambda_{f}^{(-)}(k')}{ \epsilon_{f}(k')+
\epsilon_{f}(q_f)}\gamma_1^0 \gamma_{1\mu}(1-\gamma_1^5)\Bigr\}u_b(q_b)
u_q(q_q),\end{eqnarray}
where the superscripts ``(1)" and ``(2)" correspond to Figs.~\ref{d1} and
\ref{d2},  ${\bf k}={\bf p}_f-{\bf\Delta};\
{\bf k}'={\bf q}_b+{\bf\Delta};\ {\bf\Delta}={\bf
p}_F-{\bf p}_{B_c}$;
$$\Lambda^{(-)}(p)=\frac{\epsilon(p)-\bigl( m\gamma
^0+\gamma^0({\bm{ \gamma}{\bf p}})\bigr)}{ 2\epsilon (p)}.$$
Here \cite{f} 
\begin{eqnarray*} 
p_{f,q}&=&\epsilon_{f,q}(p)\frac{p_F}{M_F}
\pm\sum_{i=1}^3 n^{(i)}(p_F)p^i,\\
q_{b,q}&=&\epsilon_{b,q}(q)\frac{p_{B}}{M_{B}} \pm \sum_{i=1}^3 n^{(i)}
(p_{B})q^i,\end{eqnarray*}
and $n^{(i)}$ are three four-vectors given by
$$ n^{(i)\mu}(p)=\left\{ \frac{p^i}{M},\ \delta_{ij}+
\frac{p^ip^j}{M(E+M)}\right\}, \quad E=\sqrt{{\bf p}^2+M^2}.$$

It is important to note that the wave functions entering the weak current
matrix element (\ref{mxet}) are not in the rest frame in general. For example, 
in the $B$ meson rest frame (${\bf p}_{B}=0$), the final  meson
is moving with the recoil momentum ${\bf \Delta}$. The wave function
of the moving  meson $\Psi_{F\,{\bf\Delta}}$ is connected 
with the  wave function in the rest frame 
$\Psi_{F\,{\bf 0}}\equiv \Psi_F$ by the transformation \cite{f}
\begin{equation}
\label{wig}
\Psi_{F\,{\bf\Delta}}({\bf
p})=D_f^{1/2}(R_{L_{\bf\Delta}}^W)D_q^{1/2}(R_{L_{
\bf\Delta}}^W)\Psi_{F\,{\bf 0}}({\bf p}),
\end{equation}
where $R^W$ is the Wigner rotation, $L_{\bf\Delta}$ is the Lorentz boost
from the meson rest frame to a moving one, and   
the rotation matrix $D^{1/2}(R)$ in spinor representation is given by
\begin{equation}\label{d12}
{1 \ \ \,0\choose 0 \ \ \,1}D^{1/2}_{q,c}(R^W_{L_{\bf\Delta}})=
S^{-1}({\bf p}_{q,c})S({\bf\Delta})S({\bf p}),
\end{equation}
where
$$
S({\bf p})=\sqrt{\frac{\epsilon(p)+m}{2m}}\left(1+\frac{\bm{\alpha}{\bf p}}
{\epsilon(p)+m}\right)
$$
is the usual Lorentz transformation matrix of the four-spinor.

The general structure of the current matrix element (\ref{mxet}) is
rather complicated, because it is necessary to integrate both with
respect to $d^3p$ and $d^3q$. The $\delta$-function in the expression
(\ref{gamma1}) for the vertex function $\Gamma^{(1)}$ permits to perform
one of these integrations. As a result the contribution of
$\Gamma^{(1)}$ to the current matrix element has the usual structure of
an overlap integral of meson wave functions and
can be calculated exactly (without employing any expansion) in the
whole kinematical range, if the wave functions of the
initial and final mesons are known. The situation with the contribution
$\Gamma^{(2)}$ is different. Here, instead of a $\delta$-function, we have
a complicated structure, containing the potential of the $q\bar
q$-interaction in the meson. It contains also the quark energies
$\epsilon_q(p)=\sqrt{m_q^2+{\bf p}^2}$, which explicitly depend on the
the relative quark momentum ${\bf p}$. The presence of such dependence
does not permit one, in the general case, to  get rid of one
of the integrations in the contribution of $\Gamma^{(2)}$ to the
matrix element (\ref{mxet}). Therefore, it is necessary to use some 
additional considerations in order to simplify calculations. The main
idea is to expand the vertex 
function $\Gamma^{(2)}$, given by (\ref{gamma2}), in such  a way that it
will be possible to use the quasipotential equation (\ref{quas}) in order
to perform one of the integrations in the current matrix element
(\ref{mxet}) and thus express this contribution to the decay matrix
element through the usual overlap integral of meson wave
functions. The realization of this strategy differs for the cases of
heavy-to-heavy and heavy-to-light transitions.

\section{Semileptonic $B$ meson decays to $D$ mesons}
\label{sec:bdsl}
For the description of semileptonic $B$ decays to ground state $D$
mesons (heavy-to-heavy transitions) it is convenient to use the HQET
parameterization for the decay matrix elements \cite{n}:
\begin{eqnarray}\label{ff}
 {\langle D(v')| \bar c\gamma^\mu b |B(v)\rangle  
\over\sqrt{M_{D}M_B}}
  &=& h_+ (v+v')^\mu + h_- (v-v')^\mu , \cr\cr
  \langle D(v')| \bar c\gamma^\mu b \gamma_5 |B(v)\rangle 
  &=& 0, \cr\cr
  {\langle D^*(v',\epsilon)| \bar c\gamma^\mu b |B(v)\rangle  
\over\sqrt{M_{D^*}M_B}}
  &=& i h_V \varepsilon^{\mu\alpha\beta\gamma} 
  \epsilon^*_\alpha v'_\beta v_\gamma ,\cr\cr
  {\langle D^*(v',\epsilon)| \bar c\gamma^\mu\gamma_5 b |B(v)\rangle  
\over\sqrt{M_{D^*}M_B}}
  &=& h_{A_1}(w+1) \epsilon^{* \mu} 
   -(h_{A_2} v^\mu + h_{A_3} v'^\mu) (\epsilon^*\cdot v) ,
   \end{eqnarray}
where $v~(v')$ is the four-velocity of $B~(D^{(*)})$ meson,
$\epsilon^\mu$ is the polarization vector of the final vector meson,
and the form
factors   $h_i$ are dimensionless functions of the product of
four-velocities $$w=v\cdot
v'=\frac{M_B^2+M_{D^{(*)}}^2-q^2}{2M_BM_{D^{(*)}}},$$
and $q=p_B-p_{D^{(*)}}$ is the momentum transfer from the parent to
daughter meson. 

In HQET these form factors up to $1/m_Q$ are given by \cite{n}
\begin{eqnarray}\label{cffgs}
h_{+}\!\!\!&=&\!\!\!\xi+(\varepsilon_c+\varepsilon_b)\left[2\chi_1-4(w-1)\chi_2+
12\chi_3\right],\cr\cr
h_{-}\!\!\!&=&\!\!\!(\varepsilon_c-\varepsilon_b)\left[2\xi_3-\bar\Lambda\xi\right],\cr\cr 
h_V\!\!\!&=&\!\!\!\xi+\varepsilon_c\left[2\chi_1-4\chi_3+
\bar\Lambda\xi\right]+\varepsilon_b\left[2\chi_1-4(w-1)\chi_2+
12\chi_3+\bar\Lambda\xi
-2\xi_3\right],\cr\cr
h_{A_1}\!\!\!&=&\!\!\!\xi+\varepsilon_c\left[2\chi_1-4\chi_3+
\frac{w-1}{w+1}\bar\Lambda\xi\right]+\varepsilon_b\left[2\chi_1-4(w-1)\chi_2+
12\chi_3+
\frac{w-1}{w+1}\left(\bar\Lambda\xi-2\xi_3\right)\right],\cr\cr
h_{A_2}\!\!\!&=&\!\!\!\varepsilon_c\left[4\chi_2-\frac2{w+1}\left(
\bar\Lambda\xi+\xi_3\right)\right],\cr\cr
h_{A_3}\!\!\!&=&\!\!\!\xi+\varepsilon_c\left[2\chi_1-4\chi_2-
4\chi_3+\frac{w-1}{w+1}\bar\Lambda\xi^{(n)}-\frac2{w+1}\xi_3\right]\cr\cr
&&+\varepsilon_b\left[2\chi_1-4(w-1)\chi_2+
12\chi_3+\bar\Lambda\xi-2\xi_3\right].
\end{eqnarray}

The calculation of the weak transition $B\to D^{(*)}$  matrix elements
using the heavy quark expansion shows that all model independent HQET
relations are satisfied in our model. In the limit of an infinitely heavy
quark all form factors are expressed through the Isgur-Wise function
\cite{iw} 
\begin{eqnarray}
&&h_+(w)=h_{A_1}(w)=h_{A_3}(w)=h_{V}(w)=\xi(w)\cr
&&h_{-}(w)=h_{A_2}(w)=0.
\end{eqnarray}
In our model this function is given as the overlap integral of
meson wave functions  \cite{fg}
\begin{equation} \label{iw}
  \xi(w)=\sqrt{\frac{2}{w+1}}\lim_{m_Q\to\infty} \int
  \frac{d^3p}{(2\pi)^3} \bar\Psi_{D}\left({\bf p}+
    2\epsilon_q(p)\sqrt{\frac{w-1}{w+1}} {\bf
      e_\Delta}\right)\Psi_B({\bf p}),
\end{equation} 
where $ {\bf e_\Delta}={\bf \Delta}/\sqrt{{\bf \Delta}^2}$ is the unit
vector in the direction of ${\bf \Delta}=M_D{\bf v}'-M_B{\bf v}$. In order
to fulfill the HQET relations (\ref{cffgs}) in the first order of the heavy
quark $1/m_Q$ expansion it is necessary to set
$(1-\varepsilon)(1+\kappa)=0$, which leads to the vanishing long-range
chromomagnetic interaction.  This condition is satisfied by our choice
of the anomalous chromomagnetic quark moment $\kappa=-1$. To fulfill the HQET
relations at second order in  $1/m_Q$ it is necessary to set
$\varepsilon=-1$ \cite{fg}. This gives an additional justification, based
on the heavy quark symmetry and heavy quark expansion in QCD, of the
choice of the characteristic parameters of our model.  In the infinitely
heavy quark mass limit the wave functions of initial $\Psi_B$ and
final $\Psi_D$ heavy mesons coincide. As the result the HQET
normalization condition \cite{n} $$\xi(1)=1$$ is exactly
reproduced. The first order Isgur-Wise functions are given by
\cite{fg}  
\begin{eqnarray}
\label{fo}
\xi_3(w) &=& (\bar\Lambda -m_q) \left(1+
\frac{2}{3}\frac{w-1}{w+1}\right)\xi(w), \nonumber\\
\chi_1(w)&=&\bar\Lambda\frac{w-1}{w+1}\xi(w), \nonumber\\
\chi_2(w)&=&-\frac{1}{32}\frac{\bar\Lambda}{w+1}\xi(w), \nonumber\\
\chi_3(w)&=&\frac{1}{16}\bar\Lambda\frac{w-1}{w+1}\xi(w),
\end{eqnarray}
where the HQET parameter $\bar \Lambda =M-m_Q$ is equal to the mean energy
of a light quark in a heavy meson $$\bar 
\Lambda=\langle \varepsilon_q\rangle\simeq0.56\ \mbox{GeV}.$$
The functions $\chi_1$ and $\chi_3$ explicitly satisfy normalization
conditions at the zero recoil point  \cite{luke} 
$$\chi_1(1)=\chi_3(1)=0,$$
arising from vector current conservation.

\begin{figure}
  \centering
  \includegraphics[width=10cm]{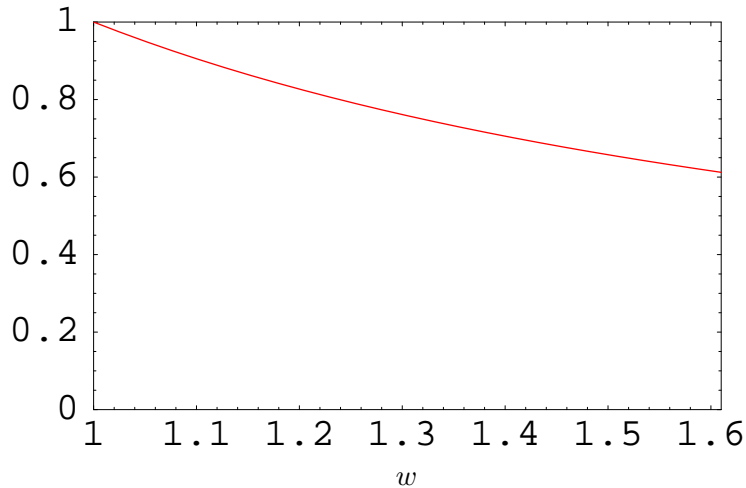}\\
\vspace*{-0.2cm}\hspace*{0.8cm}$ w$
 \caption{The Isgur-Wise function $\xi(w)$.}
  \label{fig:xi}
\end{figure}
\begin{figure}
  \centering
  \includegraphics[width=10cm]{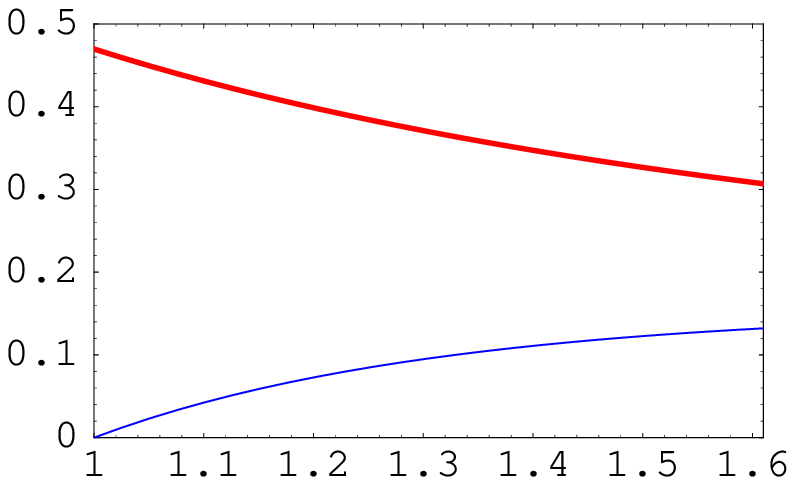}\\
\vspace*{-0.2cm}\hspace*{0.8cm}$ w$
 \caption{The first order functions $\xi_3(w)/\bar\Lambda$ (bold line)
   and $\chi_1(w)/\bar\Lambda$ (solid line).} 
  \label{fig:xi3chi1}
\end{figure}
\begin{figure}
  \centering
  \includegraphics[width=10cm]{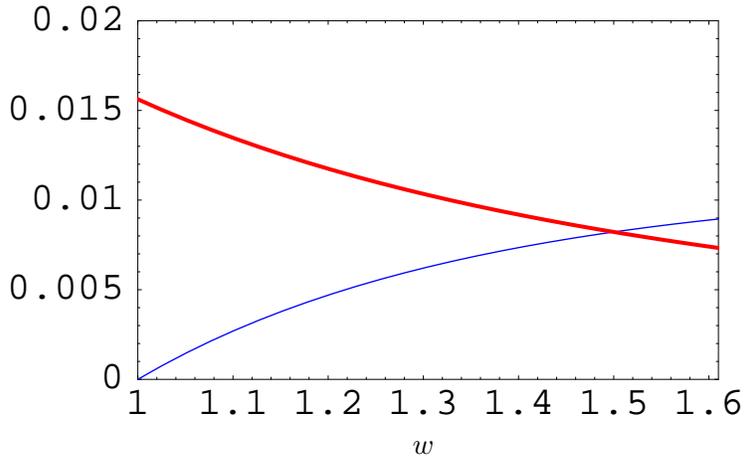}\\
\vspace*{-0.4cm}\hspace*{1.3cm}$ w$
 \caption{The first order functions $-\chi_2(w)/\bar\Lambda$ (bold
   line) and $\chi_3(w)/\bar\Lambda$ (solid line).} 
  \label{fig:chi23}
\end{figure}
In Figs.~\ref{fig:xi}--\ref{fig:chi23} we plot the Isgur-Wise
functions calculated with numerical wave functions determined in the
process of their mass calculations \cite{hlm}.  Near the zero recoil
point of the final meson $w=1$ the Isgur-Wise function can be written
as  
\begin{equation}
  \label{eq:iwexp}
  \xi(w)=1-\rho^2(w-1)+c(w-1)^2+\cdots,
\end{equation}
where  $\rho^2=-[d\xi(w)/dw]_{w=1}\simeq1.04$ is the slope and 
$c=(1/2)[d^2\xi(w)/d^2w]_{w=1}\simeq1.36$ is the curvature of the
Isgur-Wise function. The slope $\rho^2$ can be compared to the recent
quenched lattice QCD evaluation \cite{ukqcd}: $\rho^2
=0.83^{+15+24}_{-11-22}$. Note that both the slope and curvature of
the calculated Isgur-Wise function satisfy all known lower bounds (see
Ref.~\cite{lor} and references therein).

The differential semileptonic decay  rate $B\to D\,l\,\bar{\nu}$ for
the massless leptons is given by \cite{n} 
\begin{eqnarray}
\label{dgbd}
\frac{d\Gamma}{d w}=\frac{G_F^2}{48\pi^3}\,|V_{cb}|^2\,M_D^3\,
(w^2-1)^{3/2}(M_B+M_D)^2\ F_D^2(w),                                       
\end{eqnarray}
where $G_F$ is the Fermi constant, and the form factor $F_D(w)$ is defined by
\begin{eqnarray}
F_D(w)=\bigg[h_+(w)-\frac{1-r}{1+r}\,h_-(w)\bigg],  \qquad r=\frac{M_D}{M_B}.
\end{eqnarray}
Near the zero recoil point the form factor $F_D(w)$ has the following
expansion
\begin{equation}
  \label{eq:fdexp}
  F_D(w)=F_D(1)[1-\rho_D(w-1)+c_D(w-1)^2+\cdots],
\end{equation}
where the value of $F_D(1)$, calculated by using the unexpanded in
$1/m_Q$ expressions (A1)--(A3) of Ref.~ \cite{fgmslhh}, is equal to
\begin{equation}
  \label{eq:fd1}
  F_D(1)=0.966.
\end{equation}
Quenched lattice QCD calculations \cite{hkkmrs} give the value
$F_D(1)=1.058\pm0.016\pm0.03^{+0.014}_{-0.005}$, while Ref.~\cite{cln}
predicts $F_D(1)=0.98\pm0.07$.

The slope of the form factor $F_D(w)$ at zero recoil $w=1$ in our model 
\begin{equation}\label{rhod}
\rho_D^2=\left.-\frac{1}{F_D(w)}\frac{d F_D(w)}{d w}\right|_{w=1}=
0.88,
\end{equation}
is in good agreement with experimental values
\hbox{$\rho_D^2=0.76\pm0.16\pm0.08$}~\cite{cleofbd} and
\hbox{$\rho_D^2=0.69\pm0.14$}~\cite{bellefbd}, obtained by using the
linear fit of the data. The curvature of the form factor $F_D(w)$ is
equal to  
 $c_D=[1/(2F_D(1))][d^2F_D(w)/d^2w]_{w=1}\simeq0.75$.

\begin{figure}
  \centering
  \includegraphics[width=10cm,angle=-90]{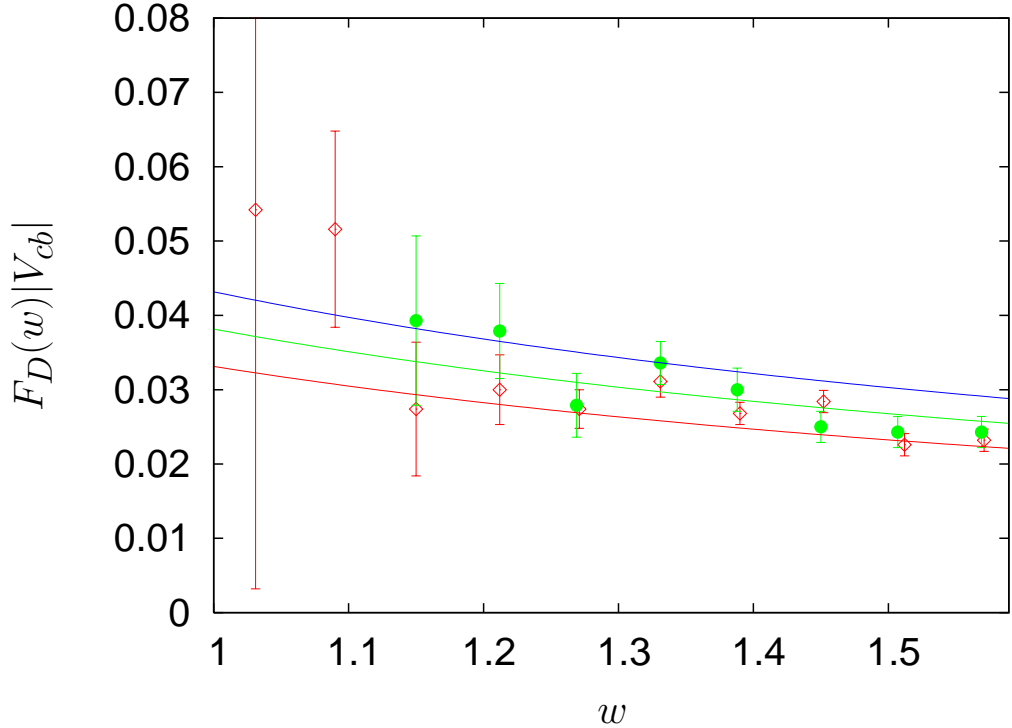}
 \caption{Comparison of experimental data and predictions of our model
   for the product 
 $F_D(w)|V_{cb}|$. Dots represent  CLEO data
   \cite{cleofbd} and diamonds  show Belle data \cite{bellefbd}. Solid lines are
   predictions of our model for 
   $|V_{cb}|=0.044,\ 0.039,\ 0.034$ (from top to bottom, respectively).}
  \label{fig:fdb}
\end{figure}

In Fig.~\ref{fig:fdb} we compare the results of our model for the
product $F_D(w)|V_{cb}|$ with the recent experimental data of CLEO
\cite{cleofbd} and  Belle \cite{bellefbd}.  It is seen that the form
factor $F_D(w)$ dependence on $w$ in our model is in good agreement
with measurements. The combined fit of CLEO 
($F_D(1)|V_{cb}|=0.039\pm0.002$) and  Belle
($F_D(1)|V_{cb}|=0.041\pm0.003$) data leads to the value of the
product of the form factor and CKM matrix element
$$F_D(1)|V_{cb}|=0.040\pm0.002.$$  
Using our prediction (\ref{eq:fd1}) for the form factor $F_D(1)$ 
we find the value of CKM matrix element 
\begin{equation}
  \label{eq:bdvcb}
  |V_{cb}|=0.0415\pm0.0020.
\end{equation}

 The differential semileptonic decay rate $B\to
D^*\,l\,\bar{\nu}$ is defined by \cite{n}
\begin{eqnarray}
\label{dgbdstar}
\frac{d\Gamma}{d w}&=&\frac{G_F^2}{48\pi^3}\,|V_{cb}|^2\,(M_B-M_{D^*})^2\,M_{D^*}^3\,
\sqrt{(w^2-1)}\,(w+1)^2\cr
&&\times\bigg[
1+\frac{4w}{w+1}\ \frac{1-2wr^*+r^{*2}}{(1-r^*)^2}
\bigg]\ F_{D^*}^2(w),       \qquad r^*=\frac{M_{D^*}}{M_B} ,                     
\end{eqnarray}
where the form factor  $F_{D^*}(w)$ is given by
\begin{eqnarray}
F_{D^*}(w)=h_{A_1}(w)\sqrt{\frac{\tilde{H}_+^2(w)+
\tilde{H}_-^2(w)+\tilde{H}_0^2(w)}
{\displaystyle 1+\frac{4w}{w+1}\ \frac{1-2wr^*+r^{*2}}{(1-r^*)^2}}} .
\end{eqnarray}
The helicity amplitudes $\tilde{H}_j(w)$ 
\begin{eqnarray}
\tilde{H}_\pm(w)&=&\frac{\sqrt{1-2wr^*+r^{*2}}}{1-r^*}\left[
1\mp\sqrt{\frac{w-1}{w+1}}\ R_1(w)
\right],\nonumber\\
\tilde{H}_0(w)&=&1+\frac{w-1}{1-r^*}\,[1-R_2(w)]
r\end{eqnarray}
are expressed through form factor ratios
\begin{eqnarray}
&&R_1(w)=\frac{h_V(w)}{h_{A_1}(w)},\nonumber\\
&&R_2(w)=\frac{h_{A_3}(w)+r^*\,h_{A_2}(w)}{h_{A_1}(w)}.
\end{eqnarray}
In the limit $m_Q\to\infty$, $R_1=R_2=1$ due to spin-flavour symmetry 
\cite{n}. Taking into account of $1/m_Q$ corrections breaks down this
symmetry relation. In our model using unexpanded in $1/m_Q$ formulas for the
form factors \cite{fgmslhh} we get the following expressions near the zero 
recoil point  $w=1$  
 \begin{eqnarray}
  \label{eq:ffexp}
  &&h_{A_1}(w)=0.918[1-0.86(w-1)+0.72(w-1)^2+\cdots],\cr
&&F_{D^*}(w)=0.918[1-0.66(w-1)+0.17(w-1)^2+\cdots],\cr
&&R_1(w)=1.39-0.23(w-1)+0.21(w-1)^2+\cdots,\cr
&&R_2(w)=0.92+0.12(w-1)-0.07(w-1)^2+\cdots.
\end{eqnarray}
It is necessary to note that the behaviour of the form factor $h_{A_1}(w)$
predicted by our model agrees, in general, with the parameterization  \cite{bgl} in
the whole kinematical range. On the other hand, the expansion of 
the form factor ratios  $R_1$ and $R_2$ is close to the QCD sum rule
results \cite{n}. Lattice QCD calculation \cite{hkmrs} gives the value
of $h_{A_1}(1)=0.9130^{+0.0238+0.0171}_{-0.0173-0.0302}$. Our
result for the difference of the slope parameters
$$\rho_{A_1}^2-\rho_{D^*}^2=0.20$$ is in agreement with previous calculations 0.21
\cite{cln} and 0.17 \cite{gl}. Comparing Eqs. (\ref{rhod}) and (\ref{eq:ffexp}) we
find the value of the slope difference 
$$\rho_{A_1}^2-\rho_{D}^2=-0.02,$$ which coincides with the one found in
Ref.~\cite{gl}.

\begin{figure}
  \centering
  \includegraphics[width=10cm,angle=-90]{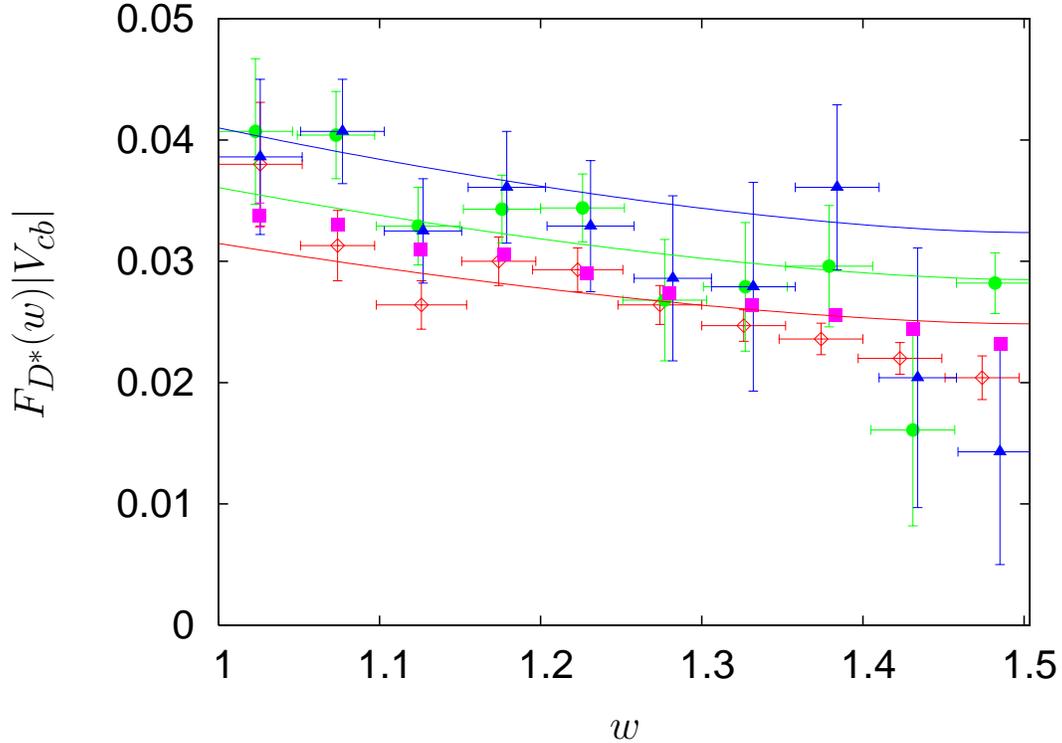}
 \caption{Comparison of experimental data and predictions of our model
   for the product $F_{D^*}(w)|V_{cb}|$. Dots show CLEO data for
   $B^+\to D^{*0} l^-\nu$, triangles -- CLEO data for $B^0\to D^{*+}
   l^-\nu$  \cite{cleofbdst}, diamonds --  Belle data 
   \cite{bellefbdst}, squares --  BaBar data \cite{babarfbdst}. Solid
   lines show predictions of our model for $|V_{cb}|=0.044,\ 0.039,\
   0.034$ (from top to bottom, respectively).} 
  \label{fig:fbdstar}
\end{figure}

\begin{table}
\caption{Comparison of our model predictions for the ratios $R_1(1)$,
  $R_2(1)$, the slope of the form factor $h_{A_1}$ for $w=1$ and
  values of the product  $F_{D^*}(1)|V_{cb}|$ with experimental data.} 
\label{tab:r1r2}
\begin{ruledtabular}
\begin{tabular}{cccccc}
  & our & CLEO~\cite{cleofbdst} & BaBar~\cite{babarfbdst} & Belle
  ~\cite{bellefbdst}& DELPHI~\cite{delphifbdst}\\ 
\hline
$R_1$ & 1.39 &1.18(30)(12) &1.396(60)(44)& & \\
$R_2$ & 0.92 &0.71(22)(7)  &0.885(40)(26) & &\\
$\rho_{h_{A_1}}^2$& 0.86 &$\left\{\begin{array}{l}0.91(15)(6)^a\cr
    1.61(9)(21)^{b}\end{array}\right.$
&$\left\{\begin{array}{l}0.79(6)^a\cr
    1.145(59)(46)^{b}\end{array}\right.$
&$\left\{\begin{array}{l}0.81(12)^a\cr
    1.35(17)^{b}\end{array}\right.$ &1.39(10)(33)$^{b}$\\
$F_{D^*}|V_{cb}|$ &0.0343(12) &$\left\{\begin{array}{l}0.0360(20)^c\cr
    0.0431(13)(18)^{b}\end{array}\right.$
&$\left\{\begin{array}{l}0.0328(5)^c\cr 
    0.0376(3)(16)^{b}\end{array}\right.$
 &$\left\{\begin{array}{l}0.0315(12)^c\cr
    0.0354(19)(18)^{b}\end{array}\right.$
&0.0377(11)(19)$^b$\\
\end{tabular}
\begin{flushleft}
${}^a$ linear fit of experimental data.\\
${}^{b}$ fit using the form factor $h_{A_1}$ parameterization \cite{bgl}.\\
${}^c$ fit using form factor predictions of our model.
\end{flushleft}
\end{ruledtabular}
\end{table}

\begin{table}
\caption{Comparison of theoretical predictions for the ratios $R_1(1)$,
  $R_2(1)$ and their derivatives $R_1'(1)$, $R_2'(1)$.} 
\label{tab:r1r2t}
\begin{ruledtabular}
\begin{tabular}{ccccc}
Ref. & $R_1(1)$ & $R_1'(1)$ & $R_2(1)$ & $R_2'(1)$\\
\hline
our       & 1.39 & $-0.23$ & 0.92 & 0.12\\
\cite{gl} & 1.25 & $-0.10$ & 0.81 & 0.11\\
\cite{cln}& 1.27 & $-0.12$ & 0.80 & 0.11\\
\cite{n}  & 1.35 & $-0.22$ & 0.79 & 0.15\\
\cite{melstech} & 1.15 &  & 0.94&  \\
\cite{ahnv}     &1.01(2)& & 1.04(1)& \\
\end{tabular}
\end{ruledtabular}
\end{table}

In Table~\ref{tab:r1r2} we give our predictions for  $R_1(1)$,
$R_2(1)$ and for the slope of the form factor $h_{A_1}$ and available
experimental data
\cite{cleofbdst,babarfbdst,bellefbdst,delphifbdst}. Our results for
$R_1$ and $R_2$  agree well with data. The calculated slope
$\rho_{h_{A_1}}^2$ is within experimental error bars for the values
obtained by using a linear fit. Comparison of our predictions for the
product $F_{D^*}(w)|V_{cb}|$ with the experimental data from CLEO
\cite{cleofbdst},  Belle \cite{bellefbdst} and BaBar \cite{babarfbdst}
is given in Fig.~\ref{fig:fbdstar}. In general there is good agreement
between the calculated form factor behaviour and available experimental
data. The values of the product $F_{D^*}(1)|V_{cb}|$, obtained by using
our form factors are compared in Table~\ref{tab:r1r2}
with the ones based on the parameterization of the form factor $h_{A_1}$
from Ref.~\cite{bgl}. Our model leads to the values
of the product $F_{D^*}(1)|V_{cb}|$ approximately 10--15\% lower, than
the values obtained using the parameterization \cite{bgl}. A combined fit of
all above mentioned experimental data in the framework of our model
gives  
$$F_{D^*}(1)|V_{cb}|=0.0343\pm0.012.$$
Using our value of  $F_{D^*}(1)=h_{A_1}(1)=0.918$ (see (\ref{eq:ffexp})) we get
\begin{equation}
  \label{eq:bdstvcb}
  |V_{cb}|=0.0375\pm0.0015.
\end{equation}
The theoretical results for the ratios $R_1(1)$,
$R_2(1)$ and their derivatives $R_1'(1)$, $R_2'(1)$ are confronted in
Table~\ref{tab:r1r2t}. In general there is a reasonable agreement both for the
form factor ratios and their slopes. Our values are very close to the
ones found using HQET and QCD sum rules \cite{n}.  

By integrating the expressions for the differential decay rates (\ref{dgbd}),
(\ref{dgbdstar}), we get predictions for the total decay rates in our model
\begin{eqnarray}
  \label{eq:gtot}
  \Gamma(B\to D l\nu)&=&9.48|V_{cb}|^2\ {\rm ps}^{-1},\cr
\Gamma(B\to D^* l\nu)&=&24.9|V_{cb}|^2\ {\rm ps}^{-1}.
\end{eqnarray}
Taking mean values of lifetimes \cite{pdg}:
$\tau_{B^0}=1.530\times 10^{-12}$~s and $\tau_{B^+}=1.671\times
10^{-12}$~s, we find
 \begin{eqnarray}
  \label{eq:brtot}
  BR(B^0\to D^+ l^-\nu)&=&14.5|V_{cb}|^2,\cr
BR(B^+\to D^0 l^+\nu)&=&15.8|V_{cb}|^2,\cr
 BR(B^0\to D^{*+} l^-\nu)&=&38.0|V_{cb}|^2,\cr
BR(B^+\to D^{*0} l^+\nu)&=&41.3|V_{cb}|^2.
\end{eqnarray}
The comparison of theoretical and experimental branching ratios
\cite{pdg} leads to the following values of the CKM matrix element
$|V_{cb}|$: 
 \begin{eqnarray}
  \label{eq:brtotexp}
  BR(B^0\to D^+ l^-\nu)^{\rm exp}=0.0214\pm0.0020 &\qquad& |V_{cb}|=0.038\pm0.002,\cr
BR(B^+\to D^0 l^+\nu)^{\rm exp}=0.0215\pm0.0022 &\qquad& |V_{cb}|=0.037\pm0.002,\cr
 BR(B^0\to D^{*+} l^-\nu)^{\rm exp}=0.0544\pm0.0023 &\qquad& |V_{cb}|=0.038\pm0.001,\cr
BR(B^+\to D^{*0} l^+\nu)^{\rm exp}=0.065\pm0.005\quad &\qquad& |V_{cb}|=0.040\pm0.002,
\end{eqnarray}
which are in good agreement with each other and with values 
(\ref{eq:bdvcb}), (\ref{eq:bdstvcb}), found from the form factor
analysis. Thus the averaged $|V_{cb}|$  over all presented experimental
measurements of semileptonic decays $B\to De\nu$ and $B\to D^*e\nu$ is
equal to  
\begin{equation}
  \label{eq:bdavvcb}
  |V_{cb}|=0.0385\pm0.0015
\end{equation}
in good agreement with  PDG \cite{pdg}
\[|V_{cb}|=0.0409\pm0.0018\quad ({\rm exclusive}). \]

\section{Semileptonic $B$ decays to light mesons}
\label{sec:blsd}

The matrix elements of weak current $J^W$ governing the weak
$B$ decays to the light pseudoscalar meson ($P=\pi$) is parameterized by
two invariant form factors. It is convenient to use the following
decomposition: 
\begin{equation}
  \label{eq:pff1}
  \langle P(p_F)|\bar q \gamma^\mu b|B(p_{B})\rangle
  =f_+(q^2)\left[p_{B}^\mu+ p_F^\mu-
\frac{M_{B}^2-M_P^2}{q^2}\ q^\mu\right]+
  f_0(q^2)\frac{M_{B}^2-M_P^2}{q^2}\ q^\mu,
\end{equation}
where $q=p_{B}-p_F$; $M_{B}$ is the $B$ meson and $M_P$ is the final
pseudoscalar meson mass.  

The corresponding matrix elements for the weak $B$ decays to the light
vector meson 
($V=\rho$) can be parameterized by four form factors: 
\begin{eqnarray}
  \label{eq:vff1}
  \langle V(p_F)|\bar q \gamma^\mu b|B(p_{B})\rangle&=
  &\frac{2iV(q^2)}{M_{B}+M_V} \epsilon^{\mu\nu\rho\sigma}\epsilon^*_\nu
  p_{B\rho} p_{F\sigma},\\ \cr
\label{eq:vff2}
\langle V(p_F)|\bar q \gamma^\mu\gamma_5 b|B(p_{B})\rangle&=&2M_V
A_0(q^2)\frac{\epsilon^*\cdot q}{q^2}\ q^\mu
 +(M_{B}+M_V)A_1(q^2)\left(\epsilon^{*\mu}-\frac{\epsilon^*\cdot
    q}{q^2}\ q^\mu\right)\cr\cr
&&-A_2(q^2)\frac{\epsilon^*\cdot q}{M_{B}+M_V}\left[p_{B}^\mu+
  p_F^\mu-\frac{M_{B}^2-M_V^2}{q^2}\ q^\mu\right], 
\end{eqnarray}
where
$M_V$ and $\epsilon_\mu$ are the mass and polarization vector of the
final vector meson. At the maximum recoil point ($q^2=0$) these form
factors satisfy the relations: 
\[f_+(0)=f_0(0),\]
\[A_0(0)=\frac{M_{B}+M_V}{2M_V}A_1(0)
-\frac{M_{B}-M_V}{2M_V}A_2(0).\]
For massless leptons form factors $f_0$ and $A_0$ do not contribute to
the semileptonic decay rates. However they give contributions to the
nonleptonic decay rates in the factorization approximation. Note that
this parameterization is completely equivalent to the one of HQET (\ref{ff}),
and corresponding form factors can be easily expressed through each other.   

In this section we calculate semileptonic decay rates of the heavy $B$
meson into light meson, $B\to \pi(\rho)e\nu$. The final meson in these
decays contains light quarks ($u$, $d$, $s$) only, thus in contrast to
decays to heavy $D$ mesons, considered previously, the application of
the expansion in inverse powers of the final active quark is not
justified. The calculation of the  contribution of the vertex function $\Gamma^{(1)}$
(\ref{gamma1}) to the decay matrix element of the weak
current (\ref{mxet}) can, as it was already in detail discussed above,
be carried out exactly, due to the presence of 
$\delta$-function, and does not require any expansion. The
calculation of the contribution $\Gamma^{(2)}$ is significantly
more difficult, since the expansion only in inverse powers of the
heavy $b$-quark mass from  the initial $B$ meson retains the dependence on
the relative momentum in the energy of the final light quark. Such
dependence does not  allow one to
perform one of the integrals in the decay matrix element
(\ref{mxet}) using the quasipotential equation. However the final light
meson has a large (compared to its mass) recoil momentum
(${\bf\Delta}\equiv{\bf p}_F-{\bf p}_B$, $|{\bf\Delta}_{\rm
  max}|=(M_{B}^2-M_F^2)/(2M_{B})\cong M_B/2\sim 2.6$~çÜ÷)  almost in
the whole kinematical range except the small region near  
$q^2=q^2_{\rm max}$ ($|{\bf\Delta}|=0$).  This also means that the
recoil momentum of the final meson is large with respect to the mean
relative quark momentum $|{\bf p}|$ in the meson  ($\sim 0.5$~GeV).
Thus one can neglect  $|{\bf p}|$ compared to $|{\bf\Delta}|$ in  the
final light quark energy
$\epsilon_{q}(p+\Delta)\equiv\sqrt{m_{q}^2+({\bf 
p}+{\bf\Delta})^2}$, replacing it by  $\epsilon_{q}(\Delta)\equiv
\sqrt{m_{q}^2+{\bf\Delta}^2}$  in expressions for the
$\Gamma^{(2)}$ contribution in accord with the large-energy expansion (see
Introduction).  This replacement removes the relative
momentum dependence in the energy of the light quark and thus permits
to perform one of the integrations in the $\Gamma^{(2)}$
contribution using the quasipotential equation. Note that the relatively
small value of this contribution, related to its proportionality to
the binding energy in the meson, and its predictable momentum dependence
allow us to extrapolate it to the whole kinematical range. The
numerical analysis (see below) shows that such extrapolation induces insignificant
uncertainties in the final results for decay rates. 

It is convenient to consider semileptonic decays $B\to(\pi,\rho) e\nu$
in the $B$ meson rest frame. Then calculating decay matrix elements it
is necessary to take into account the relativistic transformation
(\ref{wig}) of the final meson wave function from the rest frame to
the moving one with the momentum  ${\bf \Delta}$. Applying the method
described above, we find expressions for the decay matrix element and
determine the corresponding form factors. They have the following structure: 

(a) $B\to \pi$ transitions  
\begin{equation}
  \label{eq:f+}
  f_+(q^2)=f_+^{(1)}(q^2)+\varepsilon f_+^{S(2)}(q^2)
+(1-\varepsilon) f_+^{V(2)}(q^2),
\end{equation}
\begin{equation}
  \label{eq:f0}
  f_0(q^2)=f_0^{(1)}(q^2)+\varepsilon f_0^{S(2)}(q^2)
+(1-\varepsilon) f_0^{V(2)}(q^2),
\end{equation}

(b) $B\to \rho$ transitions
\begin{equation}
  \label{eq:V}
  V(q^2)=V^{(1)}(q^2)+\varepsilon V^{S(2)}(q^2)
+(1-\varepsilon) V^{V(2)}(q^2),
\end{equation}
\begin{equation}
  \label{eq:A1}
  A_1(q^2)=A_1^{(1)}(q^2)+\varepsilon A_1^{S(2)}(q^2)
+(1-\varepsilon) A_1^{V(2)}(q^2),
\end{equation}
\begin{equation}
  \label{eq:A2}
  A_2(q^2)=A_2^{(1)}(q^2)+\varepsilon A_2^{S(2)}(q^2)
+(1-\varepsilon) A_2^{V(2)}(q^2),
\end{equation}
\begin{equation}
  \label{eq:A0}
  A_0(q^2)=A_0^{(1)}(q^2)+\varepsilon A_0^{S(2)}(q^2)
+(1-\varepsilon) A_0^{V(2)}(q^2),
\end{equation}
where expressions for  $f_{+,0}^{(1)}$, $f_{+,0}^{S,V(2)}$, $A_{0,1,2}^{(1)}$,
$A_{0,1,2}^{S,V(2)}$, $V^{(1)}$ and $V^{S,V(2)}$  are rather cumbersome
and can be found in the Appendix to Ref.~\cite{fgmslhl}.
The subscripts ``(1)'' and ``(2)'' correspond to Figs.~\ref{d1} and
\ref{d2}, $S$ and $V$ denote scalar and vector potentials of the $q\bar
q$-interaction. Let us remind that the mixing coefficient $\varepsilon$ of vector
and scalar confining potentials is equal to $-1$  in our model.
Note that form factors (\ref{eq:f+})--(\ref{eq:A0}) in the
limit of the infinitely heavy $b$ quark mass and large recoil of the final light
meson explicitly satisfy all symmetry relations  \cite{clopr,efgfr} imposed
by the large energy effective theory.

In order to increase the precision and reliability  of our calculations
compared to our previous consideration \cite{fgmslhl} we do not
perform a further expansion of form factors in inverse powers of the
heavy quark mass. Moreover, the $q^2$ dependence of form factors is
explicitly determined by these formulas and thus no  ad hoc
parameterization is necessary. We also use the numerical wave
functions found in the meson mass spectrum calculations \cite{hlm,lmm} instead of
trial (Gaussian) wave functions used in Ref.~\cite{fgmslhl}. To check
the precision of the extrapolation of the form factors in the region of small
recoil of the final light meson (${\bf\Delta}=0$, $q^2=q^2_{\rm max}$)
we perform the additional consideration of  form factors in this region.
In this analysis the simplifying substitution of the light quark
energy $\epsilon_q(p)=\sqrt{p^2+m_q^2}$ by the on-shell center of mass
energy $E_q=(M^2-m_Q^2+m_q^2)/(2M)$  (\ref{ee}) in the contribution of
the vertex function   
$\Gamma^{(2)}$ is used. Such replacement is valid in the small region
around the zero recoil point. As a result in this point one gets expressions for
the form factors similar to the formulas given in the Appendix of
Ref.~\cite{efgslbcb} with obvious substitutions. Such calculation
showed that the obtained values of form factors for semileptonic
decays $B\to(\pi,\rho) e\nu$ as well as their slopes at zero recoil
$q^2=q^2_{\rm  max}$ are in good agreement with the results found from
the extrapolation of Eqs.~(\ref{eq:f+})--(\ref{eq:A0}). The
deviations are less than  1\% confirming the reliability of such
extrapolation and of final results which are based on it. 
  
The semileptonic $B\to(\pi,\rho) e\nu$ decay from factors in our model can
be approximated with good accuracy by the following expressions \cite{melstech}:

(a) $F(q^2)= f_+(q^2),V(q^2),A_0(q^2)$ 
\begin{equation}
  \label{fitfv}
  F(q^2)=\frac{F(0)}{\displaystyle\left(1-\frac{q^2}{\tilde M^2}\right)
    \left(1-\sigma_1 
      \frac{q^2}{M_{B^*}^2}+ \sigma_2\frac{q^4}{M_{B^*}^4}\right)},
\end{equation}

(b) $F(q^2)= A_1(q^2),A_2(q^2)$
\begin{equation}
  \label{fita12}
  F(q^2)=\frac{F(0)}{\displaystyle \left(1-\sigma_1
      \frac{q^2}{M_{B^*}^2}+ \sigma_2\frac{q^4}{M_{B^*}^4}\right)},
\end{equation}
where $\tilde M=M_{B}$ for $A_0$ and $\tilde M=M_{B^*}$ for all other
form factors; the values  $F(0)$ and $\sigma_{1,2}$ are given in 
Table~\ref{hlff}. The difference of fitted form factors from the
calculated ones does not exceed  1\%.

In Table~\ref{compbpiff} we give a comparison of the predictions
for the form factors of semileptonic decays  $B\to(\pi,\rho) e\nu$ at
maximum recoil point $q^2=0$, calculated in our model with 
the results of light cone QCD sum rules (LCSR)
\cite{bzpi,bzrho,kmo}, the quark model (QM) \cite{melstech}, using
relativistic dispersion relations and two recent lattice QCD (LQCD)
calculations  \cite{okamoto,hpqcdhl} with dynamical light quarks. Note
that our new results for form factors at this point coincide with previous
calculations  \cite{fgmslhl} within errors caused by the Gaussian
parameterization of the heavy-light meson wave functions used in
Ref.~\cite{fgmslhl}.  In Ref.~\cite{agrs} a model independent
constraint for the product $|V_{ub}|f_+(0)=(7.2\pm1.8)\times 10^{-4}$
was obtained using the soft-collinear effective theory and
$B\to\pi\pi$ data, which for their final value of $|V_{ub}|$ yields
$f_+(0)=0.227\pm 0.047$.

\begin{table}
\caption{From factors of semileptonic decays $B\to(\pi,\rho) e\nu$ 
  calculated in our model. Form factors $f_+(q^2)$, $V(q^2)$,
  $A_0(q^2)$ are fitted by Eq.~(\ref{fitfv}), and form factors
  $A_1(q^2)$, $A_2(q^2)$ are fitted by Eq.~(\ref{fita12}).  }
\label{hlff}
\begin{ruledtabular}
\begin{tabular}{ccccccc}
   &\multicolumn{2}{c}{{$B\to\pi$}}&\multicolumn{4}{c}{{\  $B\to\rho$\
     }}\\
\cline{2-3} \cline{4-7}
& $f_+$ & $f_0$& $V$ & $A_0$ &$A_1$&$A_2$ \\
\hline
$F(0)$&0.217&0.217 &  0.295 & 0.231& 0.269 & 0.282\\
$\sigma_1$&0.378&-0.501& 0.875 &  0.796& 0.54&1.34\\
$\sigma_2$ &-0.41& -1.50&0&-0.055&0&0.21\\
\end{tabular}
\end{ruledtabular}
\end{table} 

\begin{table}
\caption{Comparison of theoretical predictions for the form factors of 
  semileptonic decays $B\to(\pi,\rho) e\nu$  at maximum
  recoil point $q^2=0$.  }
\label{compbpiff}
\begin{ruledtabular}
\begin{tabular}{cccccc}
  & $f_+(0)$ & $V(0)$ & $A_0(0)$ &$A_1(0)$&$A_2(0)$ \\
\hline
our&0.217 & 0.295 & 0.231& 0.269 & 0.282\\
LCSR \cite{bzpi,bzrho}&0.258(31)&0.323(29)&0.303(28)& 0.242(24) &
0.221(23)\\
LCSR \cite{kmo} &0.25(5) &0.32(10) & & 0.24(8)& 0.21(9)\\
QM \cite{melstech} &0.29& 0.31&0.29&0.26&0.24\\
QM \cite{fgmslhl} &0.20(2) & 0.29(3) & & 0.26(3) &0.31(3)\\
LQCD(FNAL)\cite{okamoto}& 0.23(2)&\\
LQCD(HPQCD)\cite{hpqcdhl}& 0.27(5)&\\ 
\end{tabular}
\end{ruledtabular}
\end{table} 

\begin{figure}
\centering
  \includegraphics[width=8cm]{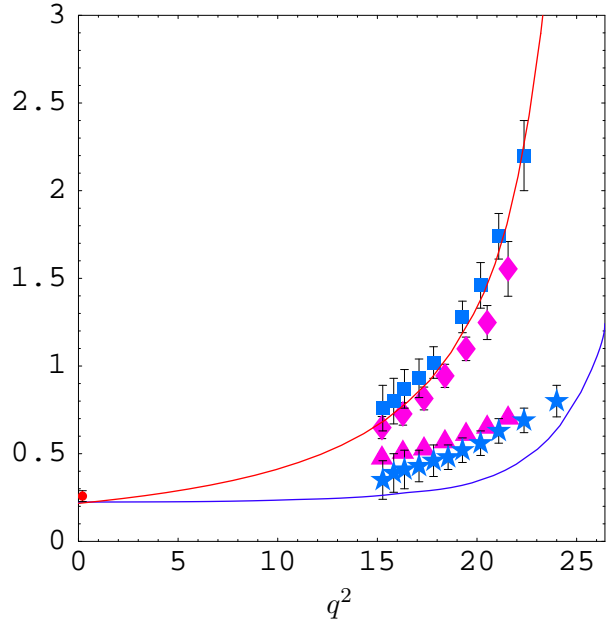}\vspace*{-0.2cm}\\
\hspace*{0.8cm} $q^2$
\caption{Form factors of the semileptonic decay $B\to\pi e\nu$  in comparison
  with lattice calculations: $f_+(q^2)$ is given by the upper plot (our), 
  squares (FNAL)\cite{okamoto}, diamonds (HPQCD)\cite{hpqcdhl};
  $f_0(q^2)$ is given by the lower plot (our), stars (FNAL)\cite{okamoto},
  triangles (HPQCD)\cite{hpqcdhl}. The LCSR value for from factors at
  $q^2=0$ is plotted by a circle \cite{bzpi}.    } 
\label{fig:fplf0}
\end{figure}

\begin{figure}
\centering
  \includegraphics[width=8cm]{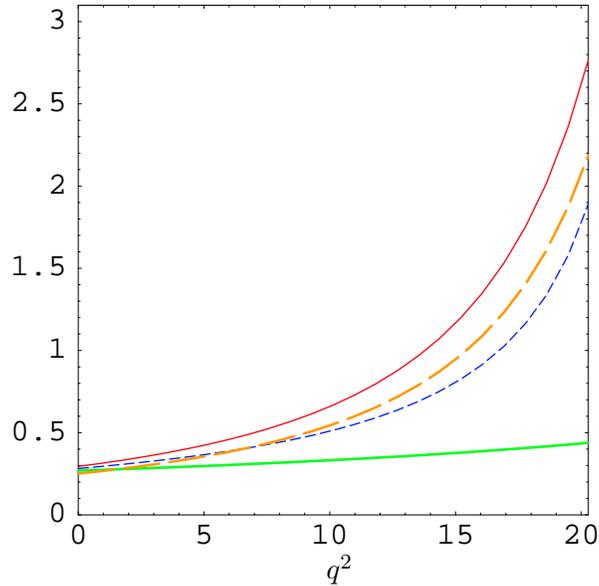}\vspace*{-0.5cm}\\
\hspace*{0.8cm} $q^2$
\caption{Form factors of the semileptonic decay $B\to\rho e\nu$: $V(q^2)$ is
  plotted by the solid line, $A_1(q^2)$ -- by the bold line,
  $A_2(q^2)$ -- by dashed line, and $A_0(q^2)$ -- by long-dashed line.     } 
\label{fig:ffrho}
\end{figure}

The  $q^2$ form factor dependence is plotted in Figs.~\ref{fig:fplf0}
and \ref{fig:ffrho}.  In Fig.~\ref{fig:fplf0} we also show recent lattice
results for the form factors  $f_+(q^2)$ and $f_0(q^2)$
\cite{okamoto,hpqcdhl}.  It is clearly seen from this figure that the
behaviour of the form factor $f_+(q^2)$  agrees with lattice computations
within errors, while our form factor $f_0(q^2)$ lies somewhat lower
than lattice data.  In this figure we also show the LCSR result
for the value of form factors at $q^2=0$ \cite{bzpi}. It agrees with
our model prediction within errors. 

We can also check the consistency of the obtained $q^2$ behaviour of the
form factor $f_+$ by comparing its calculated value at $q_{\rm max}^2$
with model independent results of chiral perturbation theory
(ChPT). For the pion recoil energy $E_\pi\sim m_\pi$ ChPT
predicts \cite{agrs}
\begin{equation}
  \label{eq:fchpt}
  f_+(q^2(E_\pi))=\frac{gf_B M_B}{2f_\pi(E_\pi+M_{B^*}-M_B)}
\left[1+O\left(\frac{E_\pi}{\Delta}\right)\right],
\end{equation}
where $g\sim 0.5$ is $B^*B\pi$ coupling \cite{agrs}, the decay
constant $f_B$ is equal 189~MeV in our model \cite{efgdc}. The
first corrections scale as $E_\pi/\Delta$, where $\Delta\sim 600$~MeV
is the mass splitting to the first radially excited $1^-$ state above
the $B^*$. Substituting these values one gets the following prediction \cite{agrs} 
\begin{equation}
  \label{eq:fpchm}
  f_+(q_{\rm max}^2)=10.38\pm 3.63,
\end{equation}
which is in  agreement with our model result $f_+(q_{\rm
  max}^2)=10.9$. 

On the other hand, the form factor $f_0$ in the soft-pion
limit $p\to 0$ and $m_\pi^2\to 0$ is related to the ratio of  
the $B$ and $\pi$ decay constants \cite{bzpi, voloshin}
\begin{equation}
  \label{eq:f0ch}
  f_0(q_{\rm max}^2)=\frac{f_B}{f_\pi}. 
\end{equation}
This relation with the above values of the decay constants gives the result
$f_0(q_{\rm max}^2)= 1.45$ again in good agreement with the prediction of
our model $f_0(q_{\rm max}^2)= 1.36$.

The differential semileptonic decay rates can be expressed in terms of
these form factors by 

(a) $B\to Pe\nu$ decay  ($P=\pi$)
\begin{equation}
  \label{eq:dgp}
  \frac{{\rm d}\Gamma}{{\rm d}q^2}(B\to Pe\nu)=\frac{G_F^2 
  \Delta^3 |V_{qb}|^2}{24\pi^3} |f_+(q^2)|^2.
\end{equation}

(b) $B\to Ve\nu$ decay ($V=\rho$)
\begin{equation}
  \label{eq:dgv}
\frac{{\rm d}\Gamma}{{\rm d}q^2}(B\to Ve\nu)=\frac{G_F^2
\Delta|V_{qb}|^2}{96\pi^3}\frac{q^2}{M_{B}^2}
\left(|H_+(q^2)|^2+|H_-(q^2)|^2   
+|H_0(q^2)|^2\right),
\end{equation}
where $G_F$ is the Fermi constant, $V_{qb}$ is CKM matrix element ($q=u$),
\[\Delta\equiv|{\bf\Delta}|=\sqrt{\frac{(M_{B}^2+M_{P,V}^2-q^2)^2}
{4M_{B}^2}-M_{P,V}^2}.
\]
Helicity amplitudes are given by the following expressions
\begin{equation}
  \label{eq:helamp}
  H_\pm(q^2)=\frac{2M_{B}\Delta}{M_{B}+M_V}\left[V(q^2)\mp
\frac{(M_{B}+M_V)^2}{2M_{B}\Delta}A_1(q^2)\right],
\end{equation}
\begin{equation}
  \label{eq:h0a}
  H_0(q^2)=\frac1{2M_V\sqrt{q^2}}\left[(M_{B}+M_V)
(M_{B}^2-M_V^2-q^2)A_1(q^2)-\frac{4M_{B}^2\Delta^2}{M_{B}
+M_V}A_2(q^2)\right].
\end{equation}

The decays rate in transversally and longitudinally polarized vector
mesons are defined by
\begin{equation}
  \label{eq:dgl}
\frac{{\rm d}\Gamma_L}{{\rm d}q^2}=\frac{G_F^2
\Delta|V_{qb}|^2}{96\pi^3}\frac{q^2}{M_{B}^2}
|H_0(q^2)|^2,  
\end{equation}
\begin{equation}
  \label{eq:dgt}
\frac{{\rm d}\Gamma_T}{{\rm d}q^2}=
\frac{{\rm d}\Gamma_+}{{\rm d}q^2}+\frac{{\rm d}\Gamma_-}{{\rm d}q^2}
=\frac{G_F^2\Delta|V_{qb}|^2}{96\pi^3}\frac{q^2}{M_{B}^2}
\left(|H_+(q^2)|^2+|H_-(q^2)|^2\right). 
\end{equation}
Integration over $q^2$ of these formulas gives the total rate
of the corresponding semileptonic decay. These rates are presented in
Table~\ref{comphlff}.  In this Table we also give the values of the partial decay
rates integrated over two intervals $q^2<16$ GeV$^2$ and
$q^2>16$ GeV$^2$ in comparison with the evaluations of LCSR in the
first interval and  LQCD in the second interval \cite{hfag}. This is
related to the fact that the predictions of these approaches are
reliable only in the above mentioned regions. The results presented in
Table~\ref{comphlff} show that our calculations agree well with the
lattice evaluations, while our values are somewhat lower than LCSR
ones which have relatively large errors. Note that our model gives the
ratio of semileptonic $B$ decay rates into $\rho$ meson with
longitudinal polarization to the corresponding rate with transverse
polarization  $\Gamma_L/\Gamma_T=0.46$.

\begin{table}
\caption{Comparison of theoretical predictions for the  rates of semileptonic
  decays $B\to(\pi,\rho) e\nu$ (in $|V_{ub}|^2$ps$^{-1}$).  }
\label{comphlff}
\begin{ruledtabular}
\begin{tabular}{cccccccccc}
& \multicolumn{4}{c}{$\Gamma$}&\multicolumn{2}{c}{{$\Gamma^{q^2<16\ {\rm GeV}^2}$}}&\multicolumn{3}{c}{{$\Gamma^{q^2>16\ {\rm GeV}^2}$
     }}\\
\cline{2-5}\cline{6-7} \cline{8-10}
Decay  & our &LCSR & FNAL &HPQCD  & our & LCSR  &our &
FNAL&HPQCD\\
 &   &  \cite{bzpi} &  \cite{okamoto} & \cite{hpqcdhl} & & \cite{bzpi}
 &  & \cite{okamoto} &  \cite{hpqcdhl}\\
\hline
$B\to\pi e\nu$ & 5.45 & 7.74(2.32)& 6.24(2.12)&6.03(1.94) & 3.68 & 5.44(1.43) & 1.77 & 1.83(50) &
1.46(35)\\
$B\to\rho e\nu$ & 13.1& & & &  10.5& & 2.60 & & \\ 
\end{tabular}
\end{ruledtabular}
\end{table}

Using mean lifetimes of $B$ mesons \cite{pdg}:
$\tau_{B^0}=1.530\times 10^{-12}$~s and $\tau_{B^+}=1.671\times
10^{-12}$~s, we find 
 \begin{eqnarray}
  \label{eq:brtot1}
  BR(B^0\to \pi^+ l^-\nu)&=&8.34|V_{ub}|^2,\cr
BR(B^+\to \pi^0 l^+\nu)&=&4.47|V_{ub}|^2,\cr
 BR(B^0\to \rho^+ l^-\nu)&=&20.1|V_{ub}|^2,\cr
BR(B^+\to \rho^0 l^+\nu)&=&10.7|V_{ub}|^2.
\end{eqnarray}
Comparison of these predictions with experimental data \cite{pdg}
leads to the following values of $|V_{ub}|$:
 \begin{eqnarray}
  \label{eq:brtotexp1}
  BR(B^0\to \pi^+ l^-\nu)^{\rm exp}=(1.36\pm0.15)\times 10^{-4} &\qquad& |V_{ub}|=(4.04\pm0.25)\times 10^{-3},\cr
BR(B^+\to \pi^0 l^+\nu)^{\rm exp}=(7.4\pm1.1)\times 10^{-4} &\qquad& |V_{ub}|=(4.07\pm0.30)\times 10^{-3},\cr
 BR(B^0\to \rho^+ l^-\nu)^{\rm exp}=(2.3\pm0.4)\times 10^{-4} &\qquad& |V_{ub}|=(3.38\pm0.30)\times 10^{-3},\cr
BR(B^+\to \rho^0 l^+\nu)^{\rm exp}=(1.24\pm0.23)\times 10^{-4} &\qquad& |V_{ub}|=(3.39\pm0.33)\times 10^{-3}.\qquad
\end{eqnarray}
Decays of the neutral and charged $B$ mesons give very close results for
$|V_{ub}|$, while the averaged values of 
$|V_{ub}|$, extracted form the decay $B\to\rho e\nu$ are approximately
16\% lower than corresponding values, found from decay $B\to\pi
e\nu$. Note that the recent CLEO \cite{picleo06} measurement of the decay branching ratio $BR(B^0\to \rho^+ l^-\nu)^{\rm exp}=(2.91\pm0.54)\times
10^{-4}$ gives  $|V_{ub}|=(3.81\pm0.35)\times 10^{-3}$ which is close
to the one extracted from the  $B\to\pi e\nu$ decay.

\begin{figure}
\includegraphics[width=7.5cm]{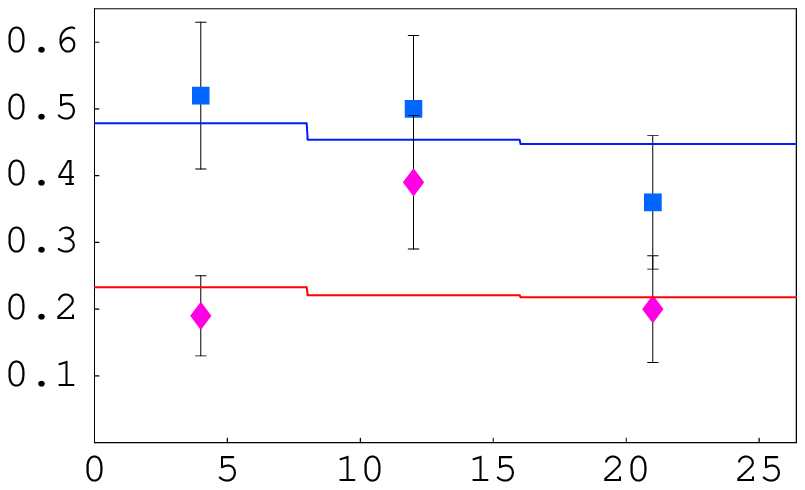} $\quad$
\includegraphics[width=7.5cm]{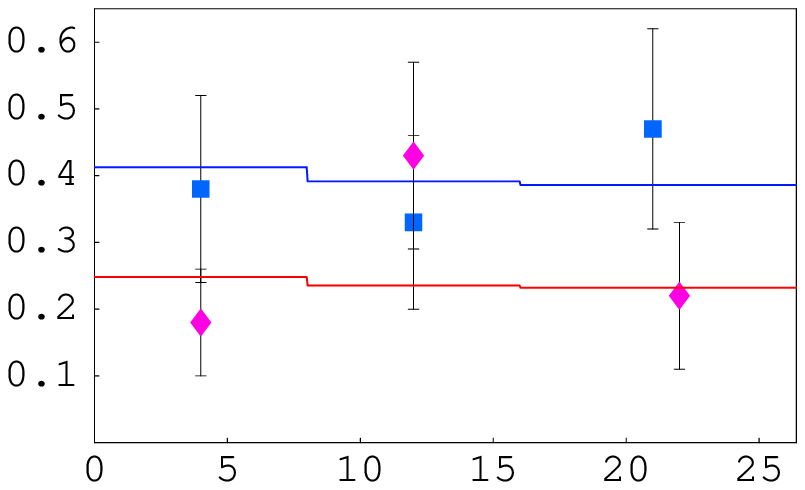}\vspace*{-0.2cm}\\
\hspace*{-3cm}\hspace*{4.cm} $q^2$ \hspace*{7.4cm} $q^2$\vspace*{-0.3cm}\\
\hspace*{-6cm}\hspace*{.2cm}(a) \hspace*{7.4cm} (b)\vspace*{0.5cm}

\includegraphics[width=7.5cm]{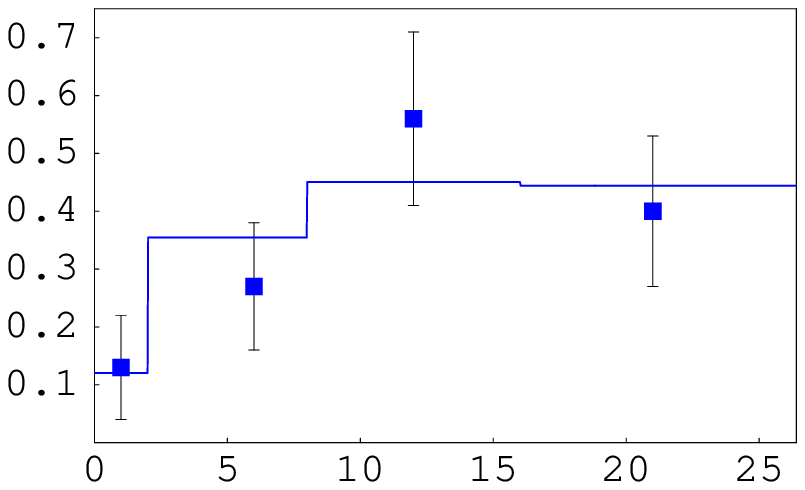} $\quad$
\includegraphics[width=7.5cm]{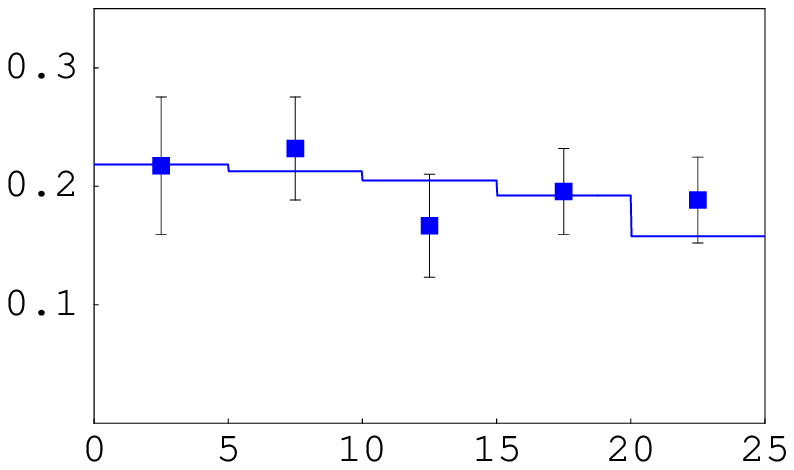}\vspace*{-0.2cm}\\
\hspace*{-3cm}\hspace*{4.cm} $q^2$ \hspace*{7.4cm} $q^2$\vspace*{-0.3cm}\\
\hspace*{-6cm}\hspace*{.2cm}(c) \hspace*{7.4cm} (d)\vspace*{0.5cm}

\includegraphics[width=7.5cm]{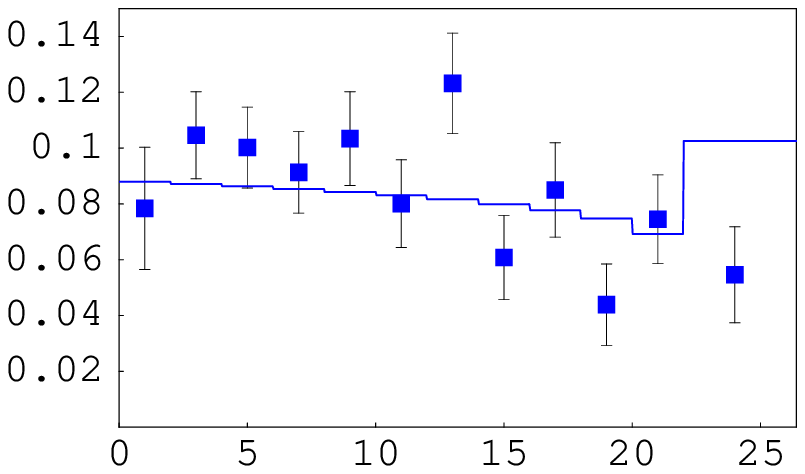}\vspace*{-0.2cm}\\
\hspace*{-7cm}\hspace*{8.2cm}$q^2$\vspace*{-0.3cm}\\
\hspace*{-10cm}\hspace*{4.5cm}(e)

\caption{Comparison of theoretical predictions for partial decay rates
  with experimental data: (a)--(c)
  $(1/\Gamma_{\rm tot})(\int_{\Delta q^2}(d\Gamma(B\to\pi
  e\nu)/dq^2)dq^2)\times 10^4$ from Refs.~\cite{pibelle06,pibabar06-2,picleo06};  (d), (e) $\Delta
  BR(B\to\pi e\nu)/BR(B\to\pi e\nu)$ from
  Refs.~\cite{pibabar05,pibabar06}, respectively.  Lower histograms on
  (a), (b) and diamonds are theoretical and experimental values for
  decays of charged $B$ meson $B^+\to\pi^0e^+\nu$. All other
  histograms and squares are theoretical predictions and experimental
  data for decays of neutral $B$ meson $B^0\to\pi^+e^-\nu$. }
\label{fig:dgammapi}
\end{figure}

\begin{figure}[tb]
\includegraphics[width=7.5cm]{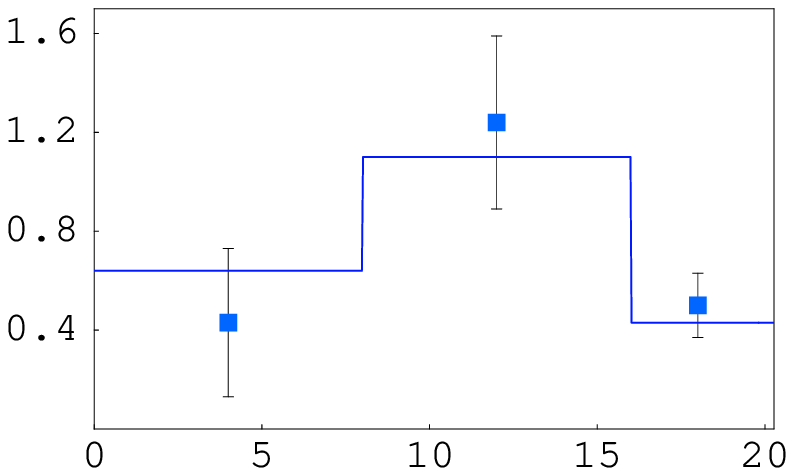} $\quad$
\includegraphics[width=7.5cm]{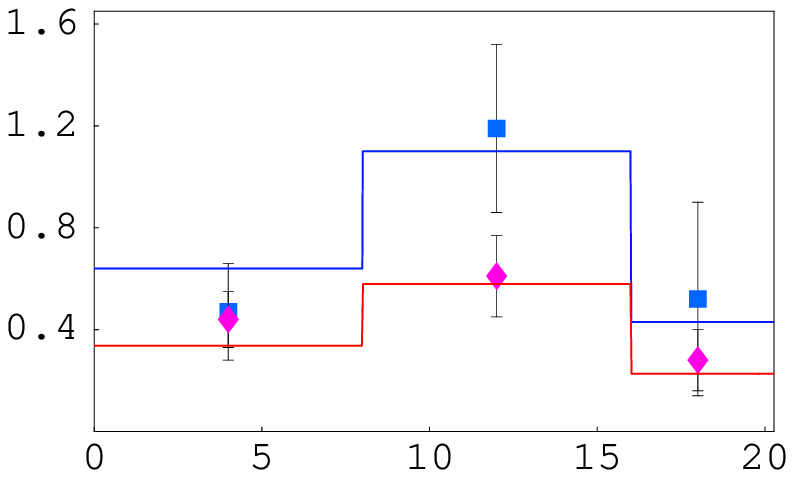}\vspace*{-0.2cm}\\
\hspace*{-3cm}\hspace*{4.cm} $q^2$ \hspace*{7.4cm} $q^2$\vspace*{-0.3cm}\\
\hspace*{-6cm}\hspace*{.2cm}(a) \hspace*{7.4cm} (b)\vspace*{0.5cm}

\includegraphics[width=7.5cm]{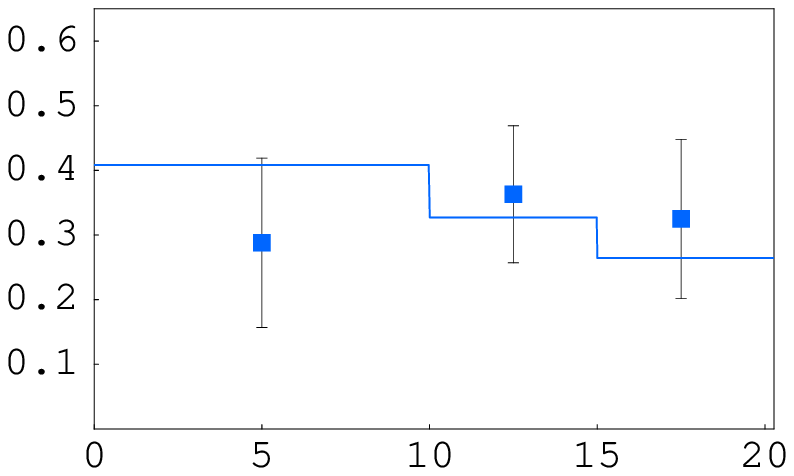}\vspace*{-0.2cm}\\
\hspace*{-7cm}\hspace*{8.2cm}$q^2$\vspace*{-0.3cm}\\
\hspace*{-10cm}\hspace*{4.5cm}(c)

\caption{Comparison of theoretical predictions for partial decay rates
  with experimental data: (a), (b)
  $(1/\Gamma_{\rm tot})(\int_{\Delta q^2}(d\Gamma(B\to\rho
  e\nu)/dq^2)dq^2)\times 10^4$ from Refs.~\cite{rhocleo03,pibelle06};  (c) $\Delta
  BR(B\to\rho e\nu)/BR(B\to\rho e\nu)$ from Ref.~\cite{pibabar05},
  respectively. Lower histograms on (b) and diamonds are theoretical
  and experimental values for 
  decays of charged $B$ meson $B^+\to\rho^0e^+\nu$.  All other
  histograms and squares are theoretical predictions and experimental
  data for decays of neutral $B$ meson $B^0\to\rho^+e^-\nu$. }
\label{fig:dgammarho}
\end{figure}

Recently significant progress has been achieved in the experimental
determination of the differential decay   $B\to(\pi,\rho) e\nu$ rate
dependence on   $q^2$.  CLEO
\cite{rhocleo03,picleo06}, BaBar
\cite{pibabar06-2,pibabar05,pibabar06}, Belle \cite{pibelle06}
measured partial decay rates in relatively narrow intervals of $q^2$. In
Figs.~\ref{fig:dgammapi} and \ref{fig:dgammarho} we present the comparison
of  our model predictions  for partial branching ratios of $B\to(\pi,\rho)
e\nu$ decays with experimental data. Plotting histograms in
Figs.~\ref{fig:dgammapi}(a)-(c) and 
Figs.~\ref{fig:dgammarho}(a),(b) we used the value of $|V_{ub}|$,
extracted from the total rate in the corresponding experiments. It is
clearly seen that our predictions agree well with almost all data for
decays of neutral as well as charged  $B$ mesons. Using these experimental
data on partial and total semileptonic $B\to(\pi,\rho) e\nu$ decay
rates it is possible to extract averaged values of  
$|V_{ub}|$:
 \begin{eqnarray}
  \label{eq:vubm}
 B\to\pi e\nu \qquad \quad &&|V_{ub}|=(4.02\pm0.10)\times 10^{-3},\cr
 B\to\rho e\nu \qquad \quad &&|V_{ub}|=(3.33\pm0.20)\times 10^{-3},
\end{eqnarray}
which are in good agreement with the ones, extracted from averaged total
decay rates (\ref{eq:brtotexp1}).

Finally, averaging over all above mentioned experimental data we get the value
of  $|V_{ub}|$ in our model 
\begin{equation}
  \label{eq:vubfinal}
  |V_{ub}|=(3.82\pm0.20)\times 10^{-3}
\end{equation}
in good agreement with PDG \cite{pdg}
\[|V_{ub}|=(3.84^{+0.67}_{-0.49})\times  10^{-3}\quad ({\rm exclusive}). \]

\section{Conclusion}
\label{sec:concl}

We calculated weak decay form factors and decay rates of
different semileptonic $B$ decays. Decays both into heavy $D^{(*)}$ and
light $\pi(\rho)$ mesons were considered. 

First, it was shown that our
relativistic quark model gives a reliable description of the
heavy-to-heavy semileptonic transitions  $B\to D^{(*)}
l\nu$. All model independent HQET relations are reproduced. The model
allows to express corresponding leading and subleading order
Isgur-Wise functions through overlap integrals of meson wave
functions. These wave functions were determined previously in the
process of the meson mass spectrum calculations. From 
the comparison with the experiment and 
predictions of other theoretical approaches it follows that our model correctly
reproduces the $q^2$ behaviour of form factors. Calculated decay
rates and branching fractions are in good agreement with data and give
very close values of the CKM matrix element $|V_{cb}|$ extracted from
different decay processes. 

Secondly, the form factors of the heavy-to-light semileptonic $B$ decays
were calculated. The consideration was done with the systematic
treatment of all relativistic effects, which are very important for
such transitions. Particular attention was paid to the inclusion of
negative-energy contributions and to the relativistic transformation
of the meson wave function from the rest to the moving reference
frame. The $q^2$ dependence of the form factors was explicitly
determined without using any ad hoc assumptions. The validity of the form
factor extrapolation, which is necessary only within the small region
near the point of zero recoil of the final light meson,  was
crosschecked by an additional 
calculation of the form factor values in this particular point. The
decay form factors are again given by the overlap integrals of the $B$
and $\pi$, $\rho$ meson wave functions, which are know from the
previous calculations of the meson mass spectra. The $q^2$ behaviour
of the form factors is in agreement with both unquenched
lattice QCD calculations and predictions of light cone QCD sum rules
within the ranges of the validity of these approaches. All this
significantly improves the reliability  of the obtained
results. The comprehensive comparison of the predictions with recent
experimental data both on total and partial decay rates allowed the
extractions of the CKM matrix element $|V_{ub}|$ which values are
rather close in different decay channels.

The evaluation of the theoretical uncertainties represents an
important problem. Unfortunately, it is not easy to estimate them in
the framework of the adopted approach. The theoretical errors within the
model, which come from  the neglected higher order terms in the $1/m_Q$
expansion for heavy-to-heavy transitions and from the form factor
extrapolation in the region of zero recoil for heavy-to-light
transitions, can be easily estimated and are less than 3\%. The main
difficulty is related to the uncertainty of the quark model
itself. However, our previous experience in calculating a vast set of different
properties of hadrons within our model indicates that such
uncertainties should not exceed 5\%. Therefore, adding these errors in
quadrature we find for the CKM matrix elements the following final
values in our model:
\begin{eqnarray}
  \label{eq:vf}
   |V_{cb}|&=&(3.85\pm0.15\pm 0.20)\times 10^{-2},\cr
  |V_{ub}|&=&(3.82\pm0.20\pm0.20)\times 10^{-3},
\end{eqnarray}
where the first error is experimental and the second one is theoretical.

\acknowledgments
The authors are grateful to M. A. Ivanov, 
M. M\"uller-Preussker and V. I. Savrin  
for support and useful discussions.  Two of us
(R.N.F. and V.O.G.)  were supported in part by the {\it Deutsche
Forschungsgemeinschaft} under contract Eb 139/2-4 and by the {\it Russian
Foundation for Basic Research} under Grant No.05-02-16243.

\end{document}